\input{aipcheck}

\documentclass[
    ,final            
  ]
  {aipproc}

\layoutstyle{6x9}

\def\Journal#1#2#3#4{{#1} {\bf #2}, #3 (#4)}

\def\NPA{{\rm Nucl. Phys.} A}

\def\PRL{\rm Phys. Rev. Lett.}

\def\PRC{{\rm Phys. Rev.} C}
\def\ZPC{{\rm Z. Phys.} C}

\def\be{\begin{equation}}
\def\ee{\end{equation}}
\def\bea{\begin{eqnarray}}
\def\eea{\end{eqnarray}}
\begin{document}

\title{Pion photoproduction on the nucleon}

\author{Haiyan Gao$^{1}$, Lingyan Zhu$^{2}$}{
  address={$^1$ Triangle Universities Nuclear Laboratory and the 
Department of Physics, Duke University, Durham, North Carolina, U.S.A.\\
$^2$ University of Illinois, Urbana-Champaign, Urbana, IL, U.S.A.\\
for the Jefferson Lab E94-104 and E02-010 Collaboration}}

\begin{abstract}
 The $\gamma n \rightarrow \pi^- p$ and 
$\gamma p \rightarrow \pi^+ n$ 
reactions are essential probes of the transition from meson-nucleon 
degrees of freedom to quark-gluon degrees of freedom in exclusive processes. 
The cross sections of these processes are also advantageous, 
for the investigation 
of oscillatory behavior around the quark counting prediction, since they 
decrease relatively slower with energy compared with other photon-induced 
processes. 
In this talk, we discuss recent results on the 
$\gamma p \rightarrow \pi^+ n$ and $\gamma n \rightarrow \pi^{-}p$ 
processes 
from Jefferson Lab experiment E94-104. 
We also discuss a new experiment in which 
singles $\gamma p \rightarrow \pi^+ n$ measurement from hydrogen, and   
coincidence $\gamma n \rightarrow \pi^{-} p$  
measurements at the quasifree kinematics from deuterium 
for center-of-mass energies between 2.3 GeV to 3.4 GeV in fine steps 
at a center-of-mass angle of $90^\circ$
are planned. 
The proposed measurement will allow a detailed investigation of the 
oscillatory scaling behavior in photopion production processes. 
\end{abstract}

\maketitle


\section{Introduction}

Exclusive processes are essential
to studies of transitions from the non-perturbative to the 
perturbative regime of Quantum Chromodynamics (QCD).
The differential cross-section for many exclusive reactions \cite{white}
at high energy and large momentum transfer appear to obey the quark counting
rule \cite{brodsky}. 
The constituent counting rule predicts the energy dependence of the
differential cross section at fixed center-of-mass angle for an
exclusive two-body reaction at high energy and large momentum transfer
as follows:

\begin{equation} 
d\sigma/dt = h(\theta_{cm})/s^{n-2},
\end{equation}

where $s$ and $t$ are the Mandelstam variables, 
$s$ is the square of the total energy in the center-of-mass frame and
$t$ is the momentum transfer squared in the $s$ channel. The quantity
$n$ is the total number of elementary fields in the initial and final states,
while $h(\theta_{cm})$ depends on details of the dynamics of the process.
The quark counting rule was originally obtained based
on dimensional analysis of typical renormalizable theories. The same rule was
later obtained in a short-distance perturbative QCD approach by 
Brodsky and Lepage\cite{lepage}.  Despite many successes, a model-independent 
test of the approach, called the hadron helicity conservation rule, tends not 
to agree with data in the similar energy and momentum region. The presence of 
helicity-violating amplitudes indicates that the short-distance expansion 
cannot be the whole story. In addition some of the cross-section data can 
also be explained in terms of non-perturbative calculations~\cite{isgur}. 

In recent years, a renewed trend has been observed in deuteron 
photo-disintegration experiments at SLAC and 
JLab \cite{gammad,schulte,rossi}. 
Onset of the scaling behavior has been 
observed in deuteron photo-disintegration \cite{gammad,schulte,rossi} 
at a surprisingly 
low momentum transfer of 1.0 (GeV/c)$^2$ to the nucleon involved. However, 
a recent polarization 
measurement on deuteron photo-disintegration \cite{krishni} 
shows disagreement with hadron helicity 
conservation in the same kinematic region where the quark counting 
behavior is apparently observed. These paradoxes make it essential to 
understand the exact mechanism governing the early onset of scaling behavior. 

The elastic proton-proton ($pp$) scattering data at
high energy and large momentum transfer have
shown very interesting characteristics: the scaled 90$^\circ$ 
center-of-mass angle data, $s^{10}{\frac{d\sigma}{dt}}$ 
show substantial oscillations about the power law behavior. 
One theoretical interpretation for such oscillatory behavior 
and the striking spin-correlation~\cite{crabb} in $pp$ scattering involves 
interference between hard pQCD short-distance and long-distance 
(Landshoff)~\cite{landshoff} amplitudes~\cite{allint} along with 
an energy dependent phase arising from 
gluonic radiative corrections. This effect is believed to be analogous to the 
Coulomb-nuclear interference that is observed in low-energy 
charged-particle scattering. Brodsky and de Teramond \cite{brodsky_de} 
suggest that the structure seen in 
$s^{10}{\frac{d\sigma}{dt}}(pp \rightarrow pp)$ and the 
large spin correlations can be attributed 
to $c\bar{c}uuduud$ resonant states. The opening of this channel gives 
rise to an amplitude with a phase shift similar to that predicted for 
gluonic radiative corrections. Deviations (oscillations) from the pQCD 
counting rule above the resonance region has recently been shown in a 
model of a composite system with two spinless charged 
constituents~\cite{close}, employing the so-called concept of
``restricted locality'' of quark-hadron duality~\cite{zhao}.  

Ji, Ma and Yuan~\cite{ji} derived a generalized counting rule for 
exclusive processes at fixed angles involving parton orbital angular momentum
and hadron helicity flip. This generalized counting rule opens 
a new window for probing the quark orbital angular momentum  
inside the nucleon.  
Recently, a non-perturbative derivation of generalized
counting rules including the orbital angular momentum was obtained by 
Brodsky and de Teramond~\cite{brodsky2003}.
Therefore, it is very important to investigate the scaling behavior 
in details and to search for QCD oscillations.
The photo-pion production reactions are well suited for these purposes
because the cross section of these processes decrease 
relatively slower with energy ($\frac{ds}{dt} \propto s^{-7}$) 
compared with other photon-induced processes. 
Rough power-law dependence of meson photoproduction seems
to agree with the constituent quark counting rule prediction
\cite{anderson76} within experimental uncertainties, for example
in the case of the $\gamma p \rightarrow \pi^{+} n$ process at a 
center-of-mass angle of $90^\circ$.
Yet it is not clear whether the counting rule 
scaling behavior has been observed in 
the $\gamma \ p \rightarrow \pi^{0}\ p$ 
process because discrepancies exist between different measurements.
For the $\gamma \ n \rightarrow \pi^{-}\ p$ process, 
no cross section data exist above a photon energy of 2.0 GeV prior to the 
recent Jefferson Lab experiment E94-104 \cite{e94104}. 

\section{Jefferson Lab Experiment E94-104}

Experiment E94-104 was carried out in Hall A~\cite{halla} at the Thomas 
Jefferson National Accelerator Facility (JLab). The continuous electron beam, 
at a current around 30 $\mu$A and energies from 1.1 to 5.5 GeV, 
impinged on a $6\%$ copper 
radiator and generated an untagged bremsstrahlung photon beam. 
The production data were taken with the 15 cm cryogenic liquid hydrogen 
(LH2) target for singles $p(\gamma,\pi^+)n$ measurement, 
or with the liquid deuterium  (LD2) target 
for coincidence $d(\gamma,\pi^-p)p$ measurement. 
The two High Resolution Spectrometers (HRS) in Hall A were used 
to detect the outgoing pions and recoil protons.
Two new aerogel \v{C}erenkov detectors in the left spectrometer 
were constructed for this experiment to provide particle 
identification for positive particles, mainly pions and protons, 
since the time-of-flight technique fails at high momentum. 
Details of the Hall A spectrometers can be found~\cite{halla}.

Based on two-body kinematics, the incident photon energy was reconstructed 
from final states, i.e. the momentum and angle of the $\pi^+$ in the 
singles measurement, momenta and angles of the $\pi^-$ and $p$ in the 
coincidence measurement. A 100 MeV bin with the center of the bin 
75 MeV from the beam energy, was chosen for the data analysis, where the 
multi-pion contribution was negligible. 
The data after background subtraction, with cuts on trigger type, 
coincidence timing, PID (particle identification), 
acceptance and photon energy, were compared to a modified 
Monte Carlo simulation code for this experiment based on MCEEP~\cite{MCEEP}
with the same cuts on acceptance and photon energy. The raw cross section 
was extracted by comparing data and simulation. 
The distributions of acceptance, reconstructed momentum 
and photon energy were in good agreement with results obtained from 
simulations. Details on the simulation and the bremsstrahlung 
photon flux calculation can be found~\cite{lyzhu_thesis}.

\begin{figure}[htbp] 
\vspace*{13pt}
\centerline{\includegraphics*[bb=14 157 535 670,width=8cm,height=5.5cm]{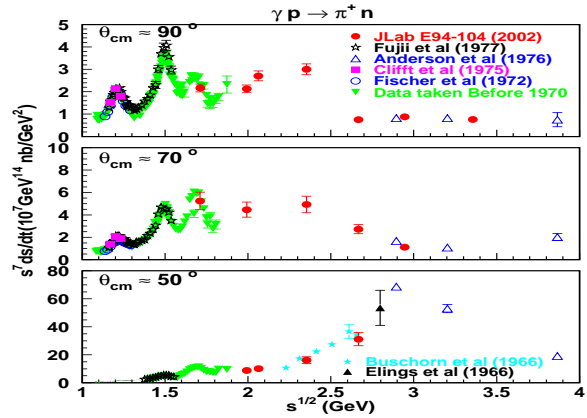}}
\vspace*{13pt}
\caption{\it The scaled differential cross section 
$s^{7}{\frac{d\sigma}{dt}}$  
versus center-of-mass energy 
for the $\gamma p \rightarrow \pi^+ n$ at 
$\theta_{cm}=90^\circ, 70^\circ, 50^\circ$. }
\end{figure}

\begin{figure}[htbp] 
\vspace*{13pt}
\centerline{\includegraphics*[bb=14 157 535 670,width=8cm,height=5.0cm]{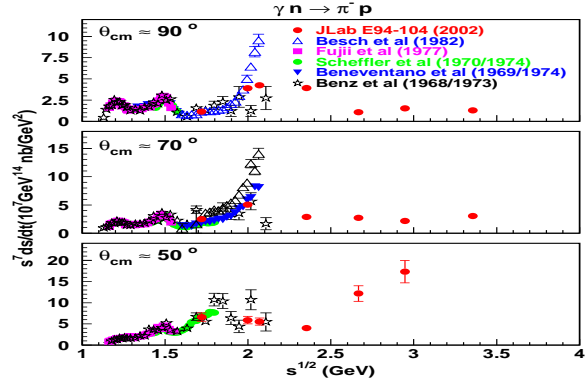}}
\vspace*{13pt}
\caption{\it The scaled differential cross section 
$s^{7}{\frac{d\sigma}{dt}}$  
versus center-of-mass energy 
for the $\gamma n \rightarrow \pi^- p$ at 
$\theta_{cm}=90^\circ, 70^\circ, 50^\circ$. }
\end{figure}  

Important correction factors were applied to deduce the final 
cross section such as the nuclear transparency factor 
in the deuteron due to the final 
state interaction, material absorption of pions and protons, pion decay loss,
detection efficiency, etc.. Analysis details can be found 
in~\cite{lyzhu_thesis,zhu}. 
The total errors were dominated by systematic errors and were estimated 
to be 10\% in cross section, while point-to-point uncertainties for the three 
kinematics at 3.3, 4.2 and 5.5 GeV to be 5\%. The statistical errors 
were approximately 2\%.
The results~\cite{zhu_longpaper}, as shown in Figure 1 and Figure 2, 
agree with the world data within uncertainties in the overlapping region. 
The data at $\theta_{cm}=70^{\circ}, 90^{\circ}$ exhibit a global 
scaling behavior predicted by the constituent counting rule 
in $\pi^{-}$ channel, similar to what was observed in the $\pi^{+}$ 
channel at similar center-of-mass angles.
The data at $\theta_{cm}=50^{\circ}$ do not display  
scaling behavior and may require higher photon energies for the 
observation of the onset of the scaling behavior. The data suggest 
that a transverse momentum
of around 1.2 GeV/c might be the scale governing the onset of scaling 
for the photo-pion production, which is consistent with 
what has been observed in deuteron photodisintegration~\cite{schulte}.
Data in these two channels at 90$^\circ$ show 
possible oscillations around the scaling behavior in similar ways 
as suggested by the insets in Fig. 3 and Fig. 4. Note that this 
possible oscillatory
behavior occurs above the known baryon resonance region.
Unfortunately, the coarse and few photon energy settings of this experiment 
do not allow us to claim the observation of oscillations.
Measurements with much finer binning, planned at JLab~\cite{dutta}, 
are essential for the confirmation of such oscillatory scaling behavior. 

\begin{figure}[htbp] 
\vspace*{13pt}
\centerline{\includegraphics*[bb=14 157 535 670,width=8cm,height=5.0cm]{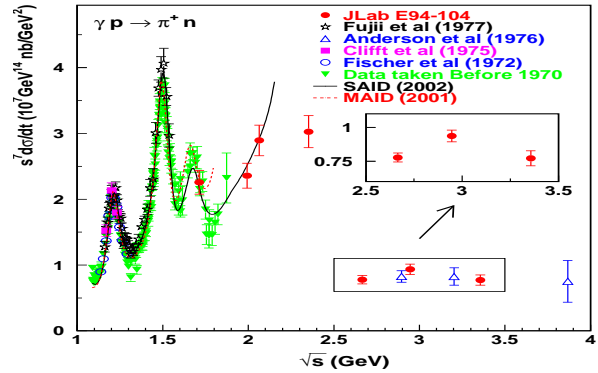}}
\vspace*{13pt}
\caption{\it The scaled differential cross section 
$s^{7}{\frac{d\sigma}{dt}}$  
versus center-of-mass energy 
for the $\gamma p \rightarrow \pi^+ n$ at 
$\theta_{cm}=90^\circ$. The data from JLab E94-104 are shown as solid circles. 
The error bars for the new data 
include statistical and systematic uncertainties, except those 
in the inset in which 
only point-to-point systematic uncertainties are included to highlight 
the possible oscillatory scaling behavior.}
\end{figure}

\begin{figure}[htbp] 
\vspace*{13pt}
\centerline{\includegraphics*[bb=14 157 535 670,width=8cm,height=5.0cm]{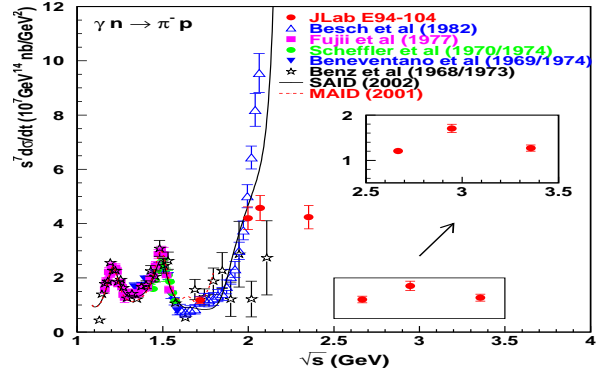}}
\vspace*{13pt}
\caption{\it The scaled differential cross section 
$s^{7}{\frac{d\sigma}{dt}}$  
versus center-of-mass energy 
for the $\gamma n \rightarrow \pi^- p$ at 
$\theta_{cm}=90^\circ$. Only point-to-point systematic uncertainties 
are shown in the inset as in Fig.3.}
\end{figure}

\section{Jefferson Lab Experiment E02-010}

Recently a new experiment \cite{dutta} was approved to carry out 
a measurement of the photo-pion production 
cross-section for the fundamental $\gamma n \rightarrow \pi^{-} p$ 
process from a $^2$H target and  for the 
$\gamma p \rightarrow \pi^{+} n$ process from a hydrogen target 
at a center-of-mass angle 
of 90$^o$, at $\sqrt{s} \sim 2.25 $ GeV to 3.41 GeV in steps 
of approximately 0.07 GeV. 
The new experiment will make individual
cross-section measurement with a 2\% statistical 
uncertainty and a point-to-point systematic uncertainty of $<$ 3\%. 
Such precision will allow the test of the oscillatory behavior 
in the scaled fundamental cross section measurement.
The proposed experiment will only be possible with the unique JLab capability 
of high luminosity and such an experiment will be carried out in Hall A at 
JLab. The projected measurements of E02-010 on the 
$\gamma n \rightarrow \pi^- p$ process at $\theta_{cm}=90^\circ$ together
with future energy upgraded 12 GeV CEBAF are shown in Fig. 5. Also shown is a 
fit of the E94-104 data based on a two-component model of 
Jain and Ralston~\cite{ralston}.

\begin{figure}[htbp] 
\vspace*{13pt}
\centerline{\includegraphics*[width=9cm,height=5.5cm]{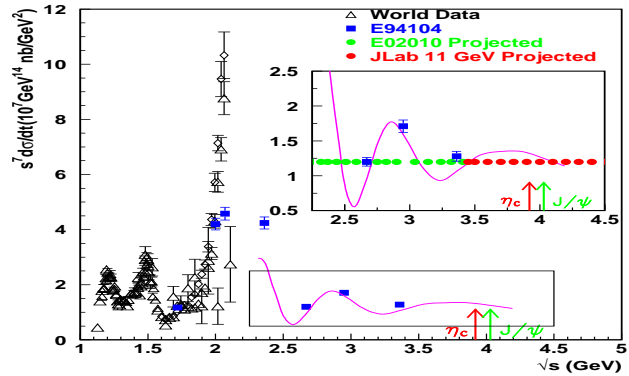}}
\vspace*{13pt}
\caption{\it The projected measurement from E02-010 and future 12 GeV energy 
upgrade at JLab on the 
scaled differential cross section 
$s^{7}{\frac{d\sigma}{dt}}$  
versus center-of-mass energy 
for the $\gamma n \rightarrow \pi^- p$ at 
$\theta_{cm}=90^\circ$.}
\end{figure}

\begin{theacknowledgments}
We thank John Arrington and Roy Holt for careful reading of this manuscript,
and Dipangkar Dutta in preparing Figure 5.
This work is supported 
by the U.S. Department of Energy under 
contract number DE-FC02-94ER40818 and DE-FG02-03ER41231.
\end{theacknowledgments}



\end{document}